\documentclass[sigconf,authorversion,nonacm]{acmart}
\usepackage{chingyi-macro}
\usepackage{enumitem}
\AtBeginDocument{%
  \providecommand\BibTeX{{
    \normalfont B\kern-0.5em{\scshape i\kern-0.25em b}\kern-0.8em\TeX}}}

\setcopyright{acmcopyright}
\copyrightyear{2024}
\acmYear{2024}
\acmDOI{10.1145/1122445.XXXXXXX}

\acmConference[CHI '24]{CHI Conference on Human Factors in Computing Systems}{May 11-May 16, 2024}{Honolulu, Hawai'i}
\acmBooktitle{CHI Conference on Human Factors in Computing Systems (CHI '24), May 11-May 16, 2024, Honolulu, Hawai'i}
\acmPrice{15.00}
\acmISBN{978-1-4503-XXXX-X/18/06}

\begin{document}

\newcommand{\projectName}{Clonemator}
\newcommand{\directSpawning}{Direct Spawning} 
\newcommand{\indirectSpawning}{Indirect Spawning} 

\newcommand{\ys}[1]{\authorcomment{GOLD}{Y}{#1}}
\newcommand{\lp}[1]{\authorcomment{BLUE}{L}{#1}}


\hidecomments
\hideoutline


\title[\projectName{}]{\projectName{}: Composing Spatiotemporal Clones to Create Interactive Automators in Virtual Reality}

\author{Yi-Shuo Lin}
\authornote{These authors contributed equally to this research.}
\email{r10922069@csie.ntu.edu.tw}
\affiliation{
  \institution{National Taiwan University}
  \streetaddress{No. 1, Sec. 4, Roosevelt Rd.}
  \city{Taipei}
  \country{Taiwan}
  \postcode{10617}
}

\author{Ching-Yi Tsai}
\authornotemark[1]
\email{ching-yi.tsai@hci.csie.ntu.edu.tw}
\affiliation{
  \institution{National Taiwan University}
  \streetaddress{No. 1, Sec. 4, Roosevelt Rd.}
  \city{Taipei}
  \country{Taiwan}
  \postcode{10617}
}

\author{Lung-Pan Cheng}
\email{lung-pan.cheng@hci.csie.ntu.edu.tw}
\orcid{0000-0002-7712-8622}
\affiliation{
  \institution{National Taiwan University}
  \streetaddress{No. 1, Sec. 4, Roosevelt Rd.}
  \city{Taipei}
  \country{Taiwan}
  \postcode{10617}
}

\renewcommand{\shortauthors}{Lin et al.}

\begin{abstract}
    \projectName{} is a virtual reality (VR) system allowing users to create their avatar clones and configure them spatially and temporally, forming automators to accomplish complex tasks. 
    In particular, clones can (1) freeze at a user's body pose as static objects, (2) synchronously mimic the user's movement, and (3) replay a sequence of the user's actions in a period of time later.    
    Combined with traditional techniques such as scaling, positional rearrangement, group selection, and duplication, \projectName{} enables users to iteratively develop customized and reusable solutions by breaking down complex tasks into a sequence of collaborations with clones.
    This bypasses implementing dedicated interaction techniques or scripts while allowing flexible interactions in VR applications. 
    We demonstrate the flexibility of \projectName{} with several examples and validate its usability and effectiveness through a preliminary user study.
    Finally, we discuss the potential of \projectName{} in VR applications such as gaming mechanisms, spatial interaction techniques, and multi-robot control and provide our insights for future research. 
\end{abstract}



\begin{teaserfigure}
\centering
\includegraphics[width=1\textwidth]{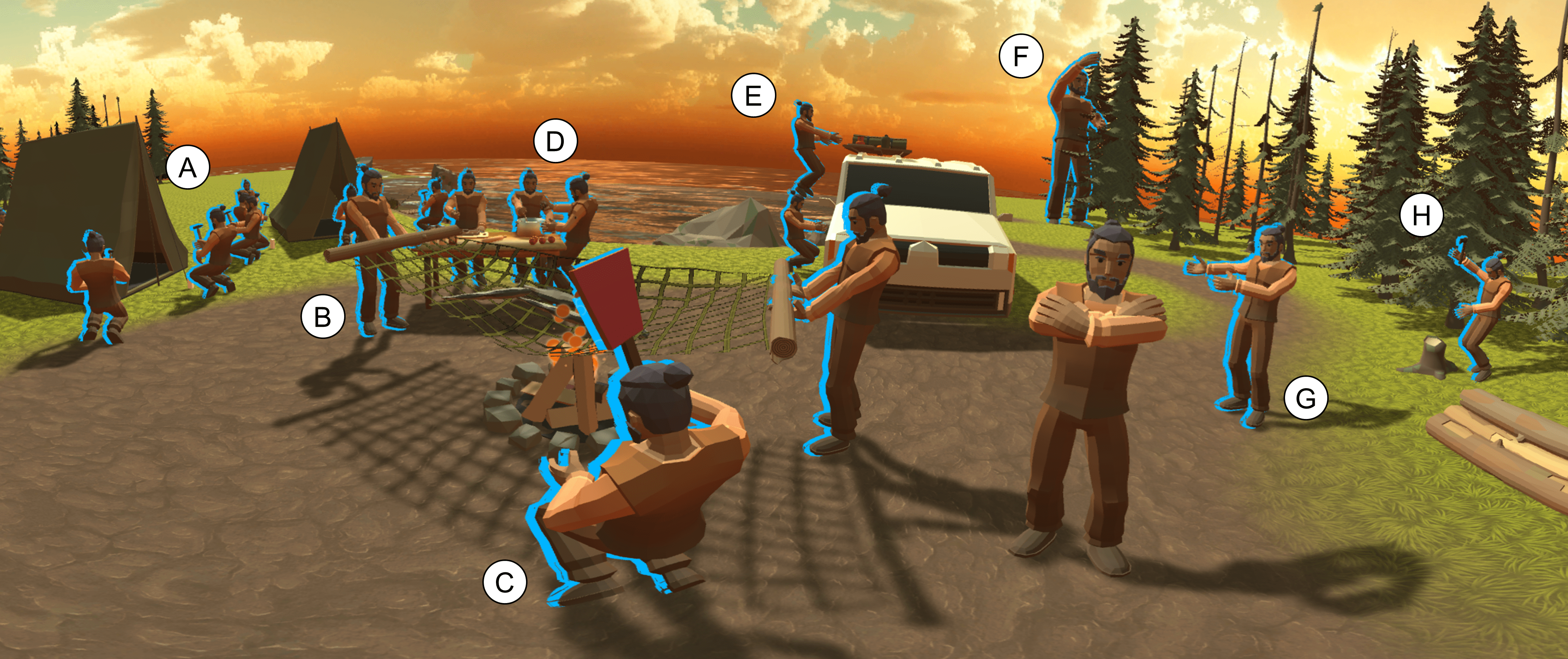}
    \caption{\projectName{} allows a user to clone their avatars in various spatiotemporal configurations and collaborate with them to achieve complex tasks in virtual reality: (A) a group of clones synchronously follow the user to hammer the tent pegs; (B) a clone mirrors the user's movement to help spread or fold the net; (C) a clone is fanning the fire while (D) the user is chopping food with a group of self-recorded clones helping cook; (E) the user steps on a clone; (F) a giant clone; (G) a body-sign clone; and (H) a remote clone replaying the logging experience previously performed by the user.}
  \label{fig:figure1}
\end{teaserfigure}

\maketitle

\section{Introduction}




Virtual reality (VR) has gained popularity with the commercialization of head-mounted displays.
Over the years, many researchers have proposed several interaction techniques to enhance experiences, including the Go-Go technique~\cite{ivan_gogointeraction_1996}, the Worlds-in-Miniature technique~\cite{Stoakley:WIM}, and portals~\cite{portals}.

In recent years, researchers have started exploring the potential of augmenting the human body in VR with duplicated body parts, such as a sixth finger~\cite{Hoyet:SixFingers}, a third arm~\cite{Drogemuller:RemappingAThirdArm}, or supernumerary hands~\cite{are4HandsBetterThan2}.
Studies even show that extra limbs can improve performance~\cite{Schjerlund:NinjaHands}.
More recently, researchers also have tried to replicate full bodies in VR, and indicated that it can affect task performance and reduce physical movements~\cite{Miura:MultiSoma}.

While each VR technique works well by itself, there is hardly any chance of combining them to create new solutions to solve complex tasks as they are usually dedicated to certain types of jobs.
Technically speaking, it is challenging to integrate all these VR techniques into a single system due to the complexity of development and unexpected exceptions.
Also, once the VR designers decide, users have little chance to adjust the settings or parameters of these VR techniques. 

In this paper, we look into finding a more general approach that enables a user to decompose complex tasks into smaller, solvable problems by combining the users themself.
Our work is analogous to previous work on automation by demonstration, such as Sikuli~\cite{yeh_sikuli_2009} and SUGILITE~\cite{li_2017_suglite} that configure automation through actions within naive users' understandability, and come up with an intuitive and enjoyable way to tackle dynamic challenges.
We present \projectName{}, a VR system that empowers users to create clones in VR.
Configuring and collaborating with spatiotemporal clones offers an understandable approach to tackling each decomposed problem, as users can utilize solutions they've used before.
\projectName{} also visually preserves spatiotemporal contexts with clones, offering a way of visualizing the entire solving procedure.



Figure~\ref{fig:figure1} depicts an overview of a user who uses \projectName{} at a campsite.
In the following section, we use this as our example walkthrough to demonstrate how the key components of \projectName{} work. 

\section{Example Walkthrough}   
\label{sec:walkthrough}
At the start of the experience, the user finds himself alone at a campsite.
His goals are to set up camp with the help of \projectName{}, including pitching a tent, preparing food, catching fish and building a campfire.

\begin{figure}[!h]
\centering
   \includegraphics[width=\columnwidth]{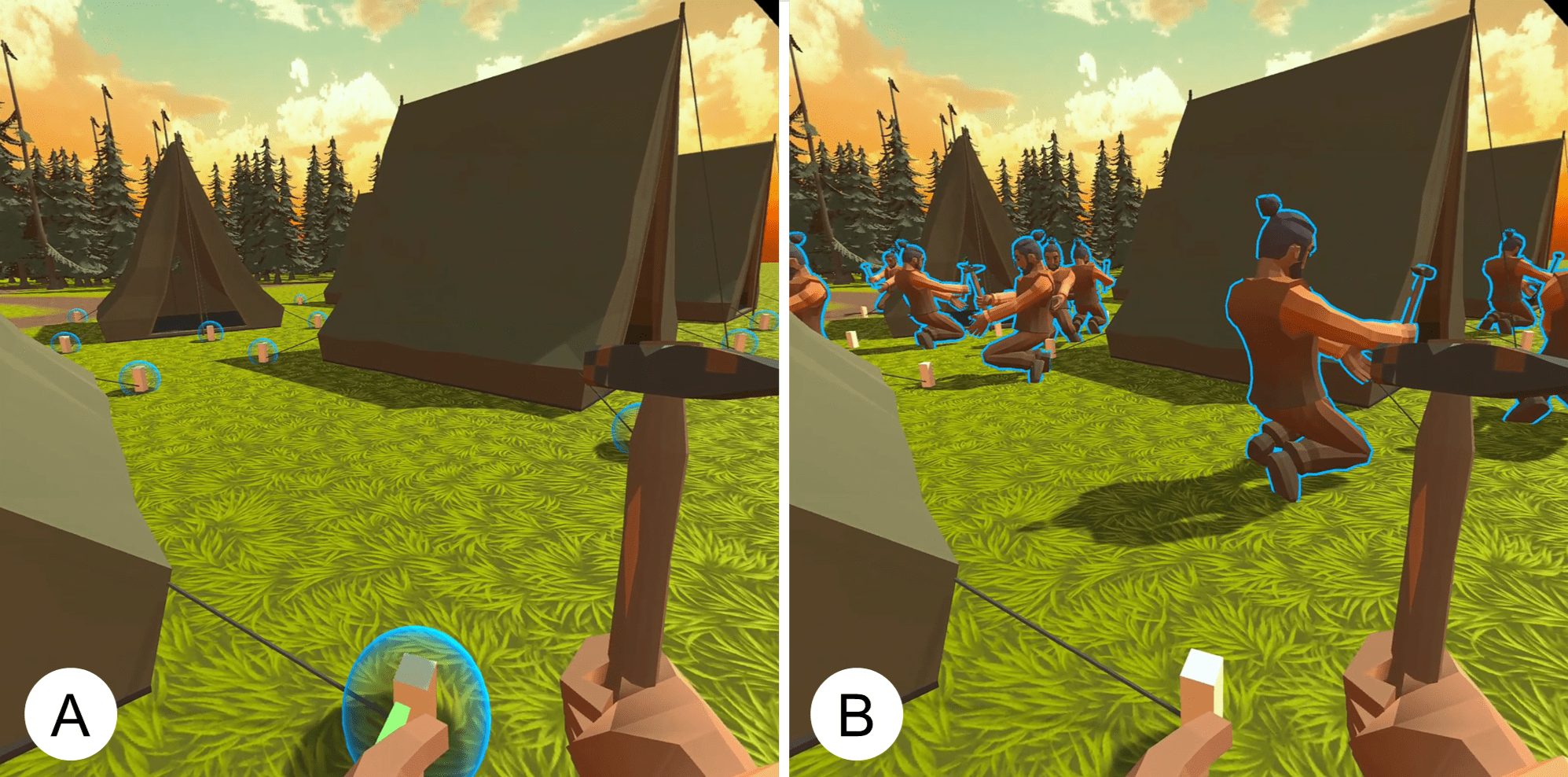}
  \caption{(A) When the user hovers on a pegs, other pegs will also be highlighted. (B) The clones will be spawned based on the offset between the user and the selected peg. Our system calculates this offset and applies it to each peg to spawn the corresponding clone, ensuring that all the clones can hammer the peg precisely.}~\label{fig:hammering}
\end{figure}

As the user walks to the tents, he notices many loose pegs surrounding the tents that need to be hammered.
The user grabs a hammer and spawns clones in front of the pegs using \textit{Auto Spawning} (Figure~\ref{fig:hammering}A).
Clones are then spawned with the same offset from the corresponding pegs, allowing the user and the clones to hammer the pegs simultaneously (Figure~\ref{fig:hammering}B).
Note that all clones are outlined in blue for easy distinction.

Next, the user decides to catch fish for dinner but finds it difficult to control the fishing net alone.
To solve this problem, he uses \textit{Relative Spawning} to spawn a clone in front of the handle on the other side of the net, ensuring that the offsets between the avatars and the handles are the same so that they can grab the two handles at the same time.
After creating a synchronous clone, the user mirrors its movement by flipping it along the vertical axis.
When the user moves backward, the clone moves backward, making spreading the fishing net easier.
On the other hand, when the user moves to the right, the clone moves to the left, allowing them to move in the same direction since they are facing each other.
To rotate the net, the user mirrors the clone's movement again to flip it back, allowing them to move synchronously and rotate in the same direction.

\begin{figure}[!h]
\centering
   \includegraphics[width=\columnwidth]{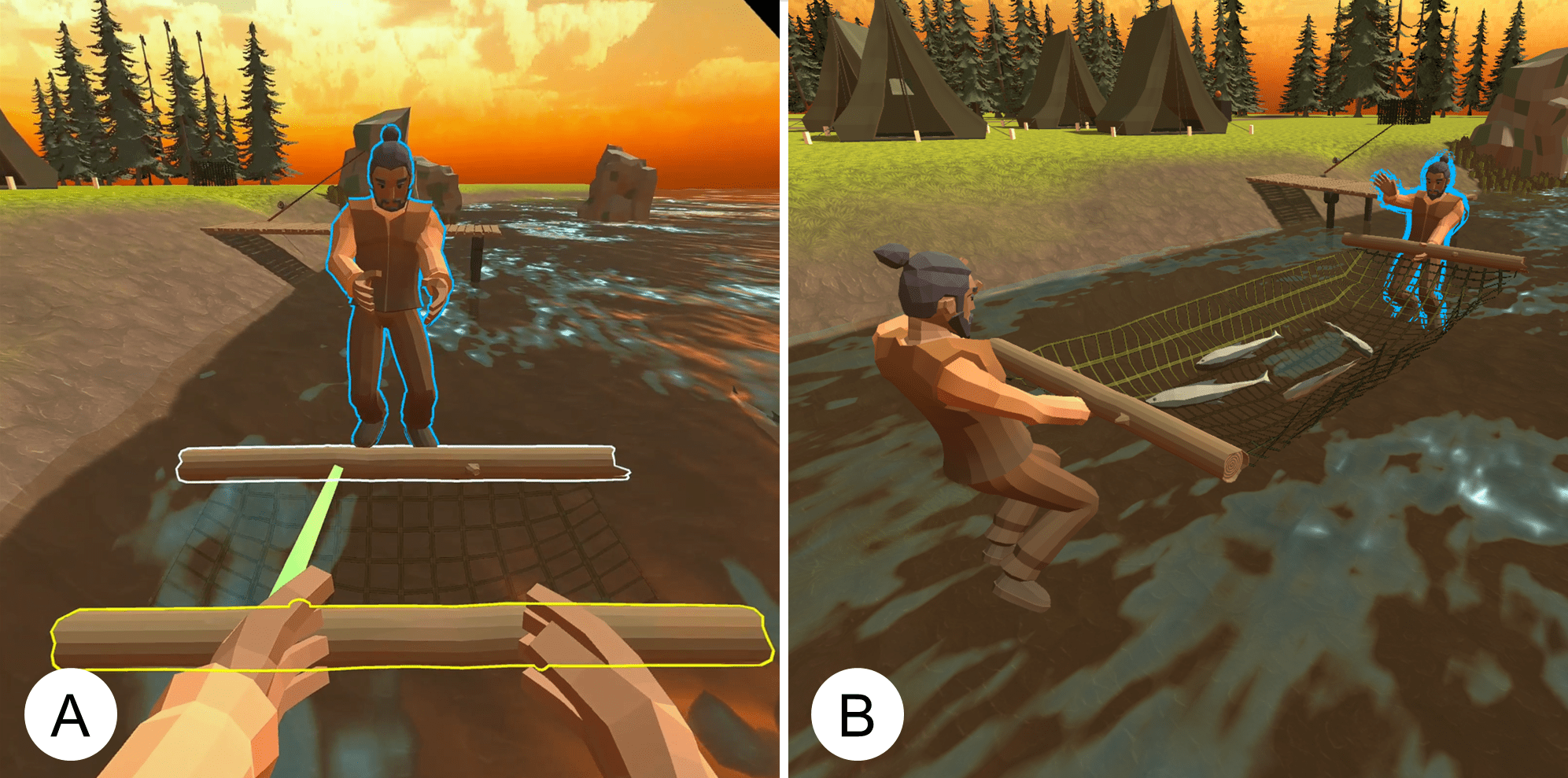}
  \caption{(A) The user spawns a clone using \textit{Relative Spawning}, which ensures that the clone can grab the handle of the other side (outlined in white) when the user grabs the handle in the front (outlined in yellow). (B) The user creates a synchronous clone and mirrors its movement, enabling them to move in the same direction and catch the fish together.}~\label{fig:catching_fish}
\end{figure}

\begin{figure}[!h]
\centering
   \includegraphics[width=\columnwidth]{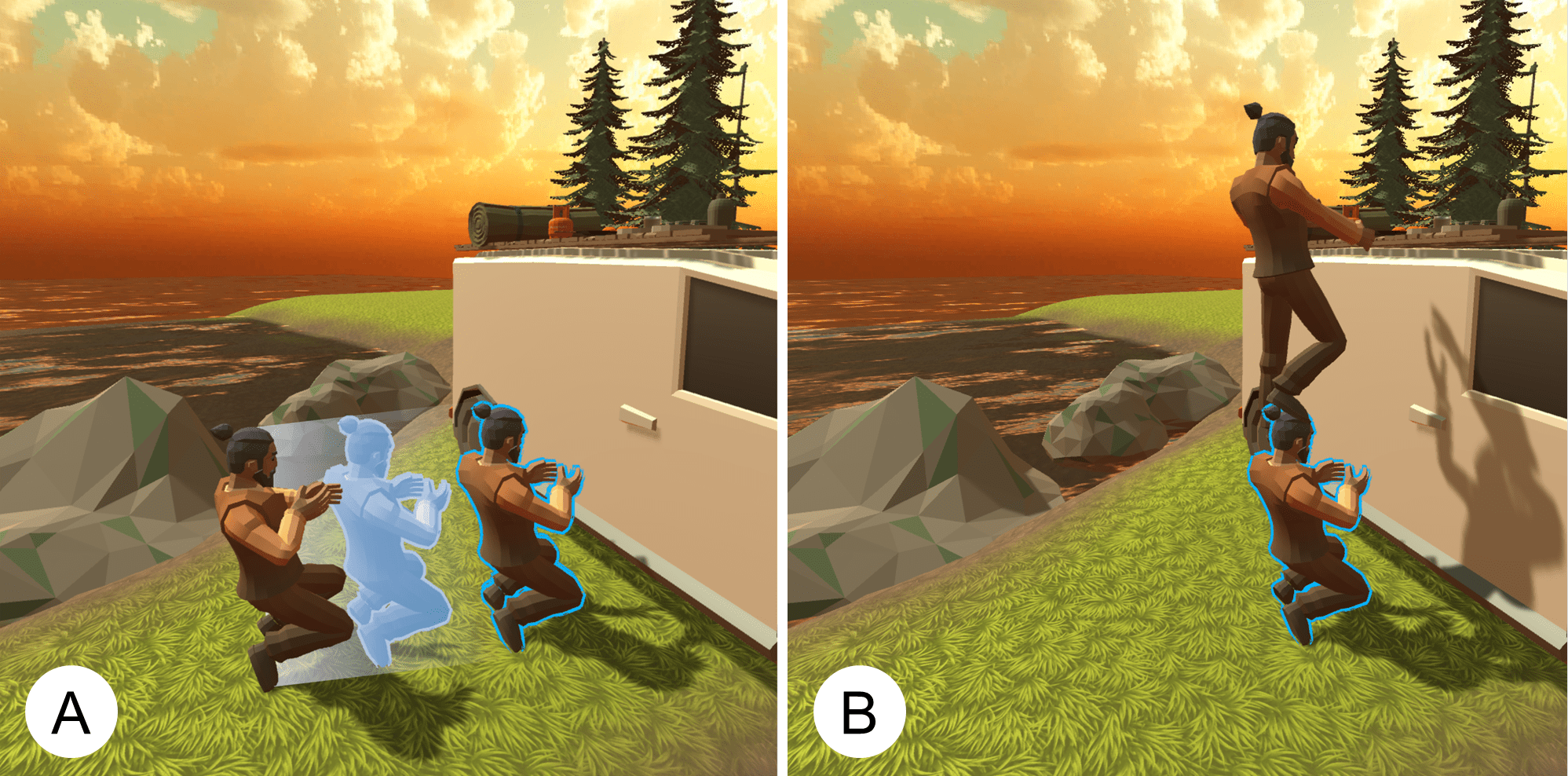}
  \caption{(A) The user crouches and performs \textit{\directSpawning{}} to create a static clone. Note that the figure has been edited to visualize the avatar's moving trajectory for easier understanding. (B) The clone now functions as a step stool, enabling the user to step onto it and reach the desired object.}~\label{fig:step_stool}
\end{figure}

Once the user catches the fish, he sets the clone to static mode and performs \textit{\directSpawning{}} to create another clone at the exact position.
The two clones each help hold one handle of the fishing net.
To move the fish and net near the table, the user can group the two clones first. 
This allows the user to directly grab one of the clones and move them together without altering the spatial relationships between the clones. 
Then, the user realizes that he needs to get a slice of beef on top of the van, which is too high to reach.
The user crouches and performs \textit{\directSpawning{}}, leaving the original body as a clone on the ground.
The clone can then serve as a stationary step stool, allowing the user to step onto it and reach the beef (Figure~\ref{fig:step_stool}). 
Note that in our current implementation, stepping onto a clone is achieved by teleporting to the clone's head since the user's legs are not tracked.

\begin{figure}[!h]
\centering
   \includegraphics[width=\columnwidth]{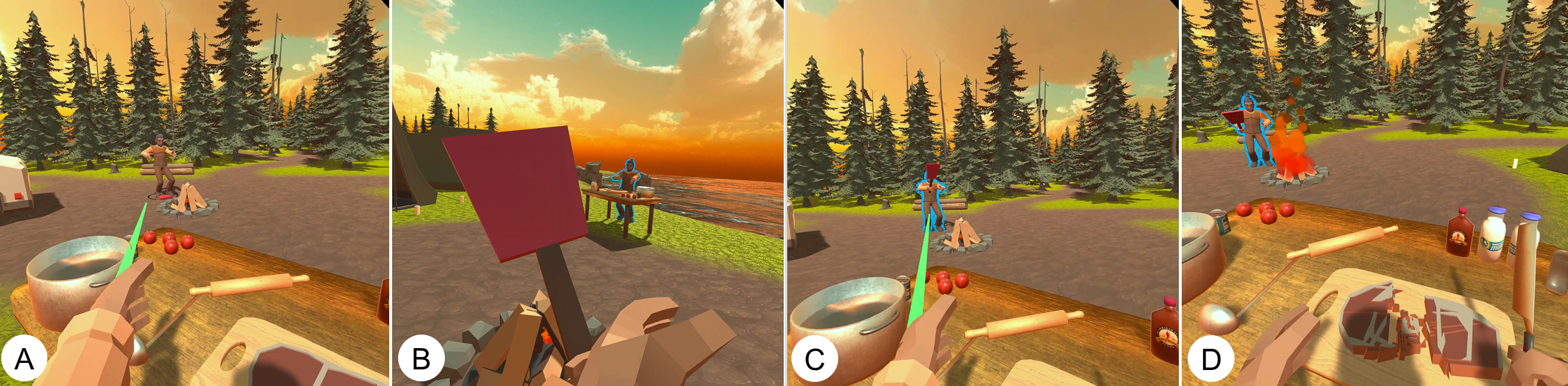}
  \caption{(A) The user spawns a clone at the desired location using \textit{\indirectSpawning{}}. (B) The user switches to the clone to grab the fan from the ground. (C) The user changes the clone's interaction mode from static to synchronous using a ray. (D) With the assistance of the clone, the user can now cut the beef while simultaneously fanning the campfire.}~\label{fig:fanning_and_cutting_beef}
\end{figure}

The user brings the beef to the table and starts cutting it with a knife but notices that the campfire is almost extinguished.
Since the user prefers not to wave his hands twice, he wants to cut the beef and fan the campfire at the same time.
The user first spawns a clone near the campfire using \textit{\indirectSpawning{}} (Figure~\ref{fig:fanning_and_cutting_beef}A) and switches to it to grab the fan (Figure~\ref{fig:fanning_and_cutting_beef}B).
The user then switches back to the original body and grabs the knife from the table.
Finally, the user updates the clone near the campfire to synchronous mode so that it can mimic the user's movement (Figure~\ref{fig:fanning_and_cutting_beef}C).
This allows the user to cut the beef and fan the campfire simultaneously since both interactions share similar movements (Figure~\ref{fig:fanning_and_cutting_beef}D).

\begin{figure}[!h]
\centering
   \includegraphics[width=\columnwidth]{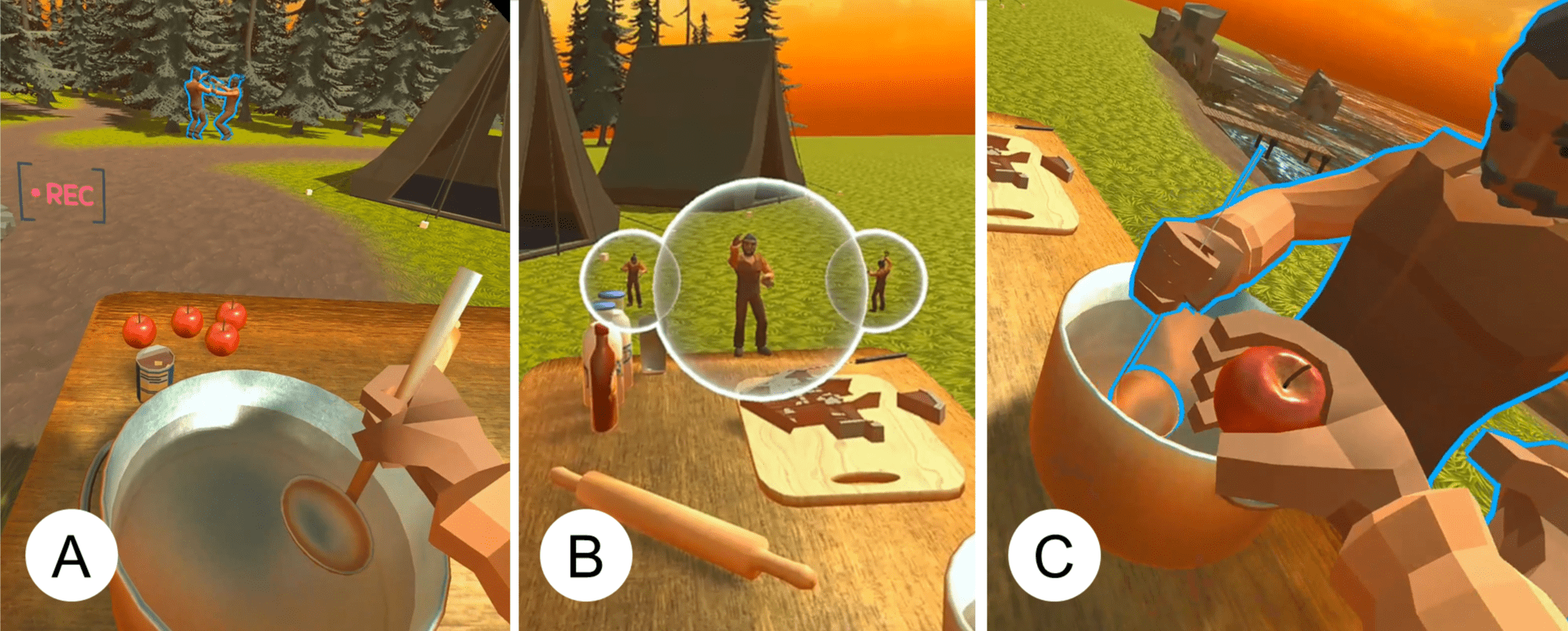}
  \caption{(A) The user records himself stirring the pot. During the recording stage, an icon appears in the upper left corner. (B) The user can preview and select previous recordings using a user interface inspired by~\cite{Xia:spacetime}. (C) Once the user applies the recorded actions to the clone, he can delegate the task of stirring the pot to it and focus on other tasks such as adding ingredients.}~\label{fig:stirring_the_soup}
\end{figure}

Then, the user wants to add ingredients such as apples and canned food to the boiling soup while keeping it stirred to prevent burning.
By recording himself stirring the pot and applying the recorded actions to a clone, the user can focus on adding ingredients while the clone continues to stir the soup (Figure~\ref{fig:stirring_the_soup}).

\begin{figure}[!h]
\centering
   \includegraphics[width=\columnwidth]{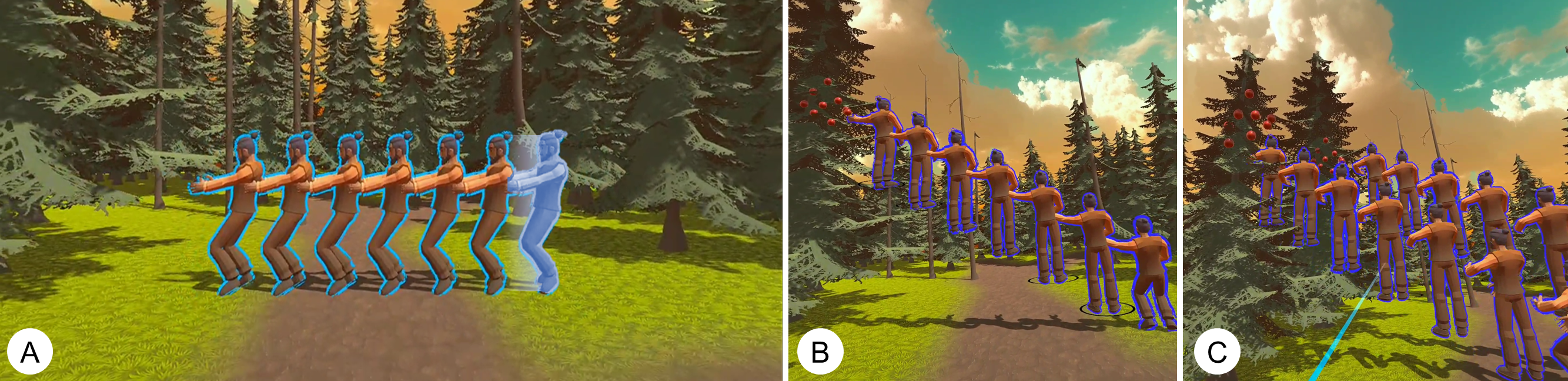}
  \caption{(A) The user performs \textit{\directSpawning{}} several times to form a chain of clones. (B) By having adjacent clones grab each other, the clones become a robotic arm that enables the user to reach apples on the tree. (C) The user can duplicate the robotic arm pattern to reuse it in other scenarios.} 
 ~\label{fig:robotic_arm}
\end{figure}

When the user wants to collect more apples, he constructs a robotic arm by arranging multiple synchronous clones in a row and having them grab each other (Figure~\ref{fig:robotic_arm}).
By adjusting the number and size of the clones, the user can achieve different control-display ratios, enabling him to customize the arm's reach and precision to suit his needs.
The flexibility of the simulated robotic arm can be further enhanced by altering the interaction modes of specific clones.
For instance, changing parts of clones to static results in a joint-like movement, allowing the user to bend and twist the robotic arm at different angles.

\begin{figure}[!h]
\centering
   \includegraphics[width=\columnwidth]{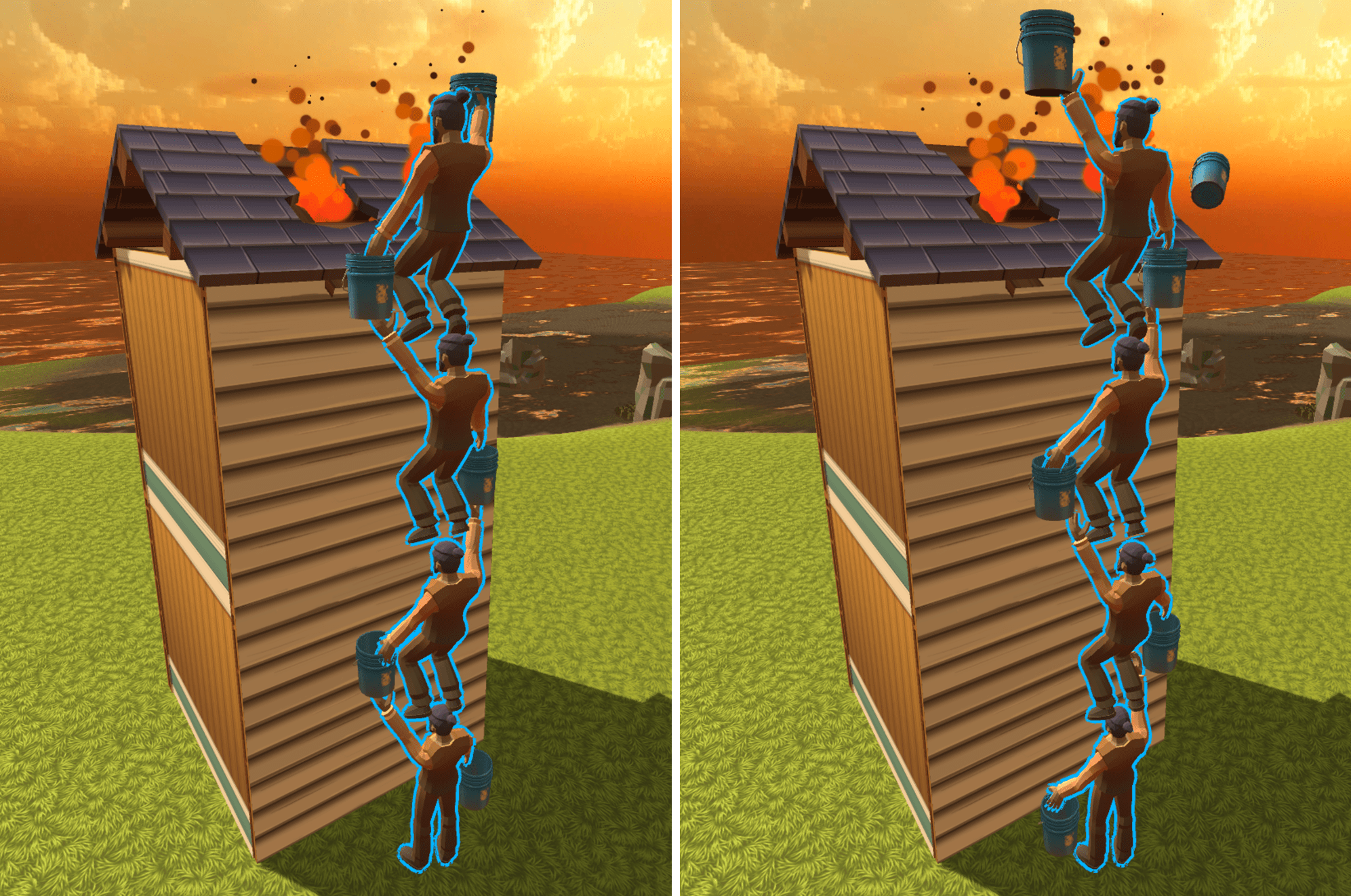}
  \caption{The user creates synchronous clones and forms a vertical bucket brigade by mirroring adjacent clones' movements. As one clone raises its left hand and lowers its right hand, the adjacent clone mirrors this movement by lowering its right hand and raising its left hand. It allows a single user to pass buckets to extinguish a fire efficiently.}~\label{fig:vertical_bucket_brigade}
\end{figure}

Suddenly, a cabin near the campsite catches fire, and the user must extinguish it quickly.
He stacks four clones vertically and sets up the clones so that the first and third clones move in sync while the second and fourth clones mirror their movements, forming a vertical bucket brigade to transfer water to the roof efficiently (Figure~\ref{fig:vertical_bucket_brigade}).

\begin{figure*}[!h]
\centering
   \includegraphics[width=\textwidth]{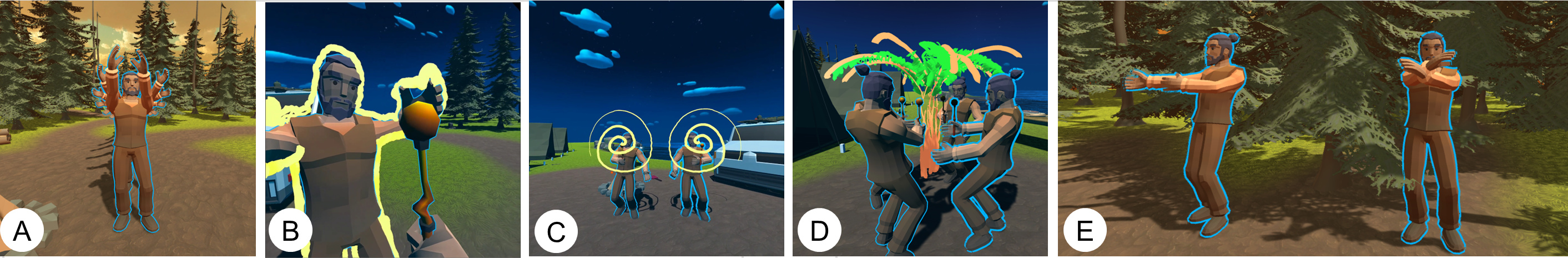}
  \caption{(A) A group of clones with delayed movements allows the user to perform the Thousand-Hand Bodhisattva Dance. When painting in VR, the user can (B) rotoscope on a clone with any postures, (C) create symmetrical paintings by mirroring the clone's movement, and (D) create symmetrical paintings by arranging the clones' positions and orientations. (E) The user can also generate customized and expressive road signs by creating clones with different postures.}~\label{fig:dancing_drawing_sign}
\end{figure*}

After extinguishing the fire, the user decides to relax by dancing.
By assigning recorded actions sequentially to multiple clones, users can delay the clones' movements and create a phase shift between them.
This allows the user to perform the Thousand-Hand Bodhisattva Dance with the clones (Figure~\ref{fig:dancing_drawing_sign}A).

When the user encounters an intersection in the forest, he leaves a static clone as a road sign to help him navigate more easily (Figure~\ref{fig:dancing_drawing_sign}E).
By creating clones that perform different postures, the user can create personalized gestures that are more intuitive than traditional flags or icons.
In addition, the user can record and replay his body movements to create even more expressive guides, allowing him to convey more detailed information.


Finally, the user takes advantage of the unique properties of clones to create painting in VR.
By using a rotoscoping technique on a clone, the user can easily sketch any desired posture (Figure~\ref{fig:dancing_drawing_sign}B).
He can also create symmetrical paintings collaboratively with synchronous clones (Figure~\ref{fig:dancing_drawing_sign}C, D).

\subsection{Summary of the Walkthrough}
\begin{figure}[!h]
\centering
    \includegraphics[width=\columnwidth]{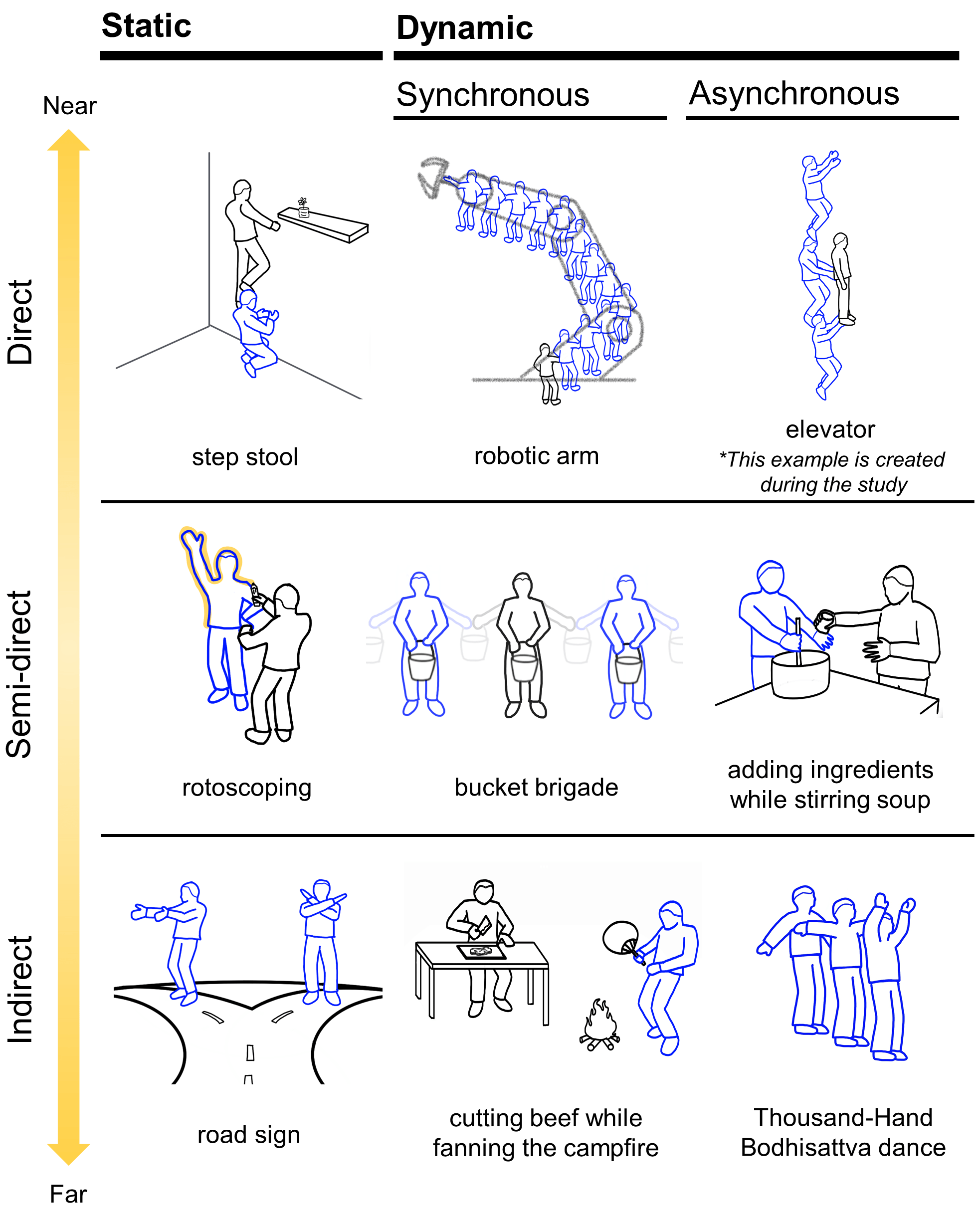}
    \caption{The solution space provided by \projectName{} and the corresponding walkthrough examples. Depending on the distance between the clones and the user, the interactions between them can be categorized as either (1) direct, (2) semi-direct, or (3) indirect. In terms of the temporal aspect, the clones can be (1) static, (2) synchronous, or (3) asynchronous. These spatial and temporal dimensions together define the possibilities of \projectName{}.}~\label{fig:taxonomy}
\end{figure}

While we demonstrate the effectiveness of~\projectName{} within a camping scenario, it is important to note that all the showcased spatiotemporal interactions are general and applicable to other scenarios and tasks as well.
To provide a comprehensive overview of the interactions enabled by \projectName{}, we summarize their taxonomy in Figure ~\ref{fig:taxonomy}. 

\section{Contributions}

The key benefit that \projectName{} brings is its capability to maintain spatial and temporal \textbf{continuity}~\cite{schlottmann1999seeing, spacetimecausality}, and thus result in the high \textbf{understandability} for the users to build functional interactions.
While there have been methods for composing VR interaction techniques, the predominant approach involves employing dedicated programming languages within a graphical user interface (GUI) and separately observing the results in 3D spatiotemporal environments. 
This discrepancy results in a discontinuity between the spatiotemporal context within VR and the process of constructing interactions (GUI coding).
In this paper, we argue that \projectName{} is the interface for maintaining seamless spatiotemporal continuity for building VR interaction techniques. 
While every interaction technique includes \textbf{users} itself and a corresponding input \text{action}, \projectName{} empowers users to temporally reuse a previous \textbf{action} and spatially configure the \textbf{users} themself (in the form of clone). 
By combining these spatial and temporal manipulations, \projectName{} enables users to build complex solutions and to adapt interaction techniques, all originating from the fundamental components of simple demonstrations of themselves, using the most familiar interaction they know.

In this paper, we make the following contributions:
\begin{itemize}[leftmargin=*]
    \item We explore the possibilities and potential for interactions between clones by providing several examples and discussing our insights.
    \item We develop a VR system that allows users to create clones with different interaction modes, including static, synchronous and replayed.
    We also integrate several additional techniques such as duplicate, mirror, and group to offer users flexibility in composing clones and building their own automators.
    \item We conduct a preliminary user study to validate the intuitiveness and effectiveness of the key components of \projectName{}.
\end{itemize}

\section{\projectName{}}
\label{sec:Clonemator}

The key design space of \projectName{} is spatial and temporal controls of clones.

In this section, we begin by outlining the essential components necessary for enabling \projectName{} and how we implement them. Then, we elaborate on how these key components can act as unifying operations for composing other existing interaction techniques for forming new, dynamic, and reusable automators or interactions. 


\subsection{Key Component \#1: Spatial Manipulations}
The core spatial manipulations facilitating \projectName{} encompass three fundamental aspects: (1) \textbf{Spawning Method} and \textbf{Duplication} for clones, (2) \textbf{Switching Control} among differently located clones, and (3) \textbf{Grouping} a configured set of clones.

\subsubsection*{\textbf{Spawning Method}}
The fundamental atomic action in \projectName{} is cloning the users themselves. In our current implementation, anticipating that the spawning method would be the most frequently used, we offer four variations of clone spawning. These four methods essentially represent the same function through different polymorphisms: Direct, Indirect, Auto, and Relative.
\begin{itemize}[leftmargin=*]
\item\textit{\directSpawning{}:} allows the user to create a clone at the exact position and in the same posture as the current avatar.
When triggering \textit{\directSpawning{}}, our system duplicates the current avatar and smoothly moves the user backward by a customized offset, simulating the feeling of leaving the original body.
The original body then becomes a clone, facilitating precise spawning.
It is useful when the user wants to preserve the context of his current interaction, such as when holding a ladle (Figure~\ref{fig:stirring_the_soup}A).
In this case, the clone will continue to hold the ladle while the user can move away and perform other actions.

\item\textit{\indirectSpawning{}:} enables users to immediately spawn a clone at the desired location using a ray.
Users can adjust the clone's rotation along the vertical axis by pressing the thumbstick to the left or right.



\item\textit{Auto Spawning} and \textit{Relative Spawning:} exhibit a higher level of context awareness.
Using \textit{Auto Spawning}, the user can select an object and spawn clones in front of other objects of the same kind while preserving the relative position and rotation offsets.
In our current implementation, we detect objects of the same kind by searching for objects with the same tag that is manually assigned beforehand.
\textit{Relative Spawning} can be used when the user wants to spawn clones in front of arbitrary objects.
When selecting a reference object, \projectName{} calculates the offset between the reference object and the user.
Then, as the user selects the next object, \projectName{} spawns a clone based on the position of the selected object and the calculated offset.
Keeping the same offset between objects ensures that the user can interact with those objects simultaneously.

When spawning a clone while holding an object using \textit{Auto Spawning} or \textit{Relative Spawning}, the object is also duplicated and distributed to the clone's hand since we aim to allow users to perform the same interactions with their clones (Figure~\ref{fig:hammering}B).
We also set the default interaction modes of the clones spawned by these methods to synchronous.
\end{itemize}



\subsubsection*{\textbf{Switch Control}}
While users can create numerous clones simultaneously, human's ability to manage multiple entities with multiple perspectives is limited~\cite{Miura:MultiSoma}. Consequently, implementing a feature that enables users to transition between clones becomes imperative.
Our current implementation allows users to switch between clones by selecting the desired clone using a ray.
After switching to a clone's body, the user can gain first-person control over its movements, enabling more accurate adjustments and fine-tuning of the clone's position and interaction.
To minimize potential disorientation after switching, we implement an interpolation method to smoothly transition the positions and rotations of the user's viewpoint between the original avatar and the clone. 
We also reduce the field of view during the transition to mitigate motion sickness. While this implementation employs one perspective for the users at the same time, an alternative approach could involve a multi-perspective system such as OVRlap~\cite{Schjerlund:OVRlap}.


\subsubsection*{\textbf{Group}}
\textit{Group} locks the spatial relationship between multiple clones.
In our current implementation, users can group multiple clones by selecting them with a ray.
Once the clones are grouped, they will be visually distinguished with the same outline color.
Users can manipulate them by moving, applying interaction modes, duplicating, and removing them.
Grouping serves to maintain the existing spatial relationships between clones.
Without grouping, every time users wish to reuse a previous automator (\ie{} a group of clones), it would potentially lead to the reconfiguration of the relationship between them, such as relative positions or rotational offset.

\subsubsection*{\textbf{Duplicate}}
The duplicate function is indispensable to facilitate the reuse of previous automators (\ie{} a group of clones), which enables users to reuse complete automators with configured properties.
In our current implementation, users can grab and pull off a single clone or a group of clones from a distance using a ray to make a duplication, which preserves the original recorded interactions or properties of the original automator.
Although the interaction of creating a duplicate through pulling off an avatar is similar to that in SpaceTime~\cite{Xia:spacetime}, the main difference in \projectName{} is that the duplicated clone can also interact with the virtual world.
Notably, duplicating a single clone is equivalent to spawning one. Therefore, an alternative implementation of \projectName{} involves retaining only the duplication function, as duplicating is inclusive of spawning. \cy{this would probably need to be moved to the spawning section.}



\subsection{Key Component \#2: Temporal Manipulations}
To enable \projectName{}, our temporal manipulations offer three interaction modes for clones: (1) \textbf{Static}, (2) \textbf{Synchronous}, and (3) \textbf{Replayed}.



\subsubsection*{\textbf{Static Mode}}
In \textit{Static Mode}, the clone freezes in time and remains stationary.
It is valuable when the user needs a stable platform or needs help holding objects in place.
For example, the user can hang a lantern on the clone to illuminate a dark forest.
This frees up the user's hands and allows the user to perform other tasks such as gathering woods. 
Additionally, a static clone can serve as a reference point.
For example, the user can leverage a static clone's body as a measuring tool for comparing the height of a door when building a house. 
While \textit{group} locks the spatial relationship between multiple clones, \textit{static} mode locks the temporal property of a  single clone.

\subsubsection*{\textbf{Synchronous Mode}}
In this mode, the clone follows the user's movements precisely, creating a one-to-one replication of the user's actions.
Users can create a group of synchronous clones to perform repetitive tasks or collaborate with them synchronously.
It is also useful for activities that require synchronization among multiple clones, such as passing objects over a long distance using the bucket brigade technique.
    
\subsubsection*{\textbf{Replayed Mode}}
\projectName{} allows a clone to perform previously recorded sequences of actions in a loop repeatedly.
This allows the users to record their physical movements and their interaction events. 
Users can access a 3D carousel menu that shows previews of all previous recordings to choose the desired one and apply it to an existing clone or the users' own avatar themself. 
This mode is useful when the user needs to repeat certain interaction in the future.  One way to conceptualize the replayed mode is as a form of temporal grouping that captures a series of past user actions and interactions. When a sequence of movements is recorded and replayed, the time offset between these movements remains fixed.
Moreover, it allows users to collaborate to achieve more complex interactions with their past selves.
Currently, users can set a clone's interaction mode in our implementation by pointing a ray.

These modes represent various temporal behaviors for clones: Static mode keeps the clone fixed at a single moment, Synchronous mode allows the clone to perform future actions alongside the user, and Replayed mode enables the clone to reenact past actions.
\subsection{Composing Automators through Spatiotemporal Manipulation}
\begin{figure*}[!h]
\centering
   \includegraphics[width=\textwidth]{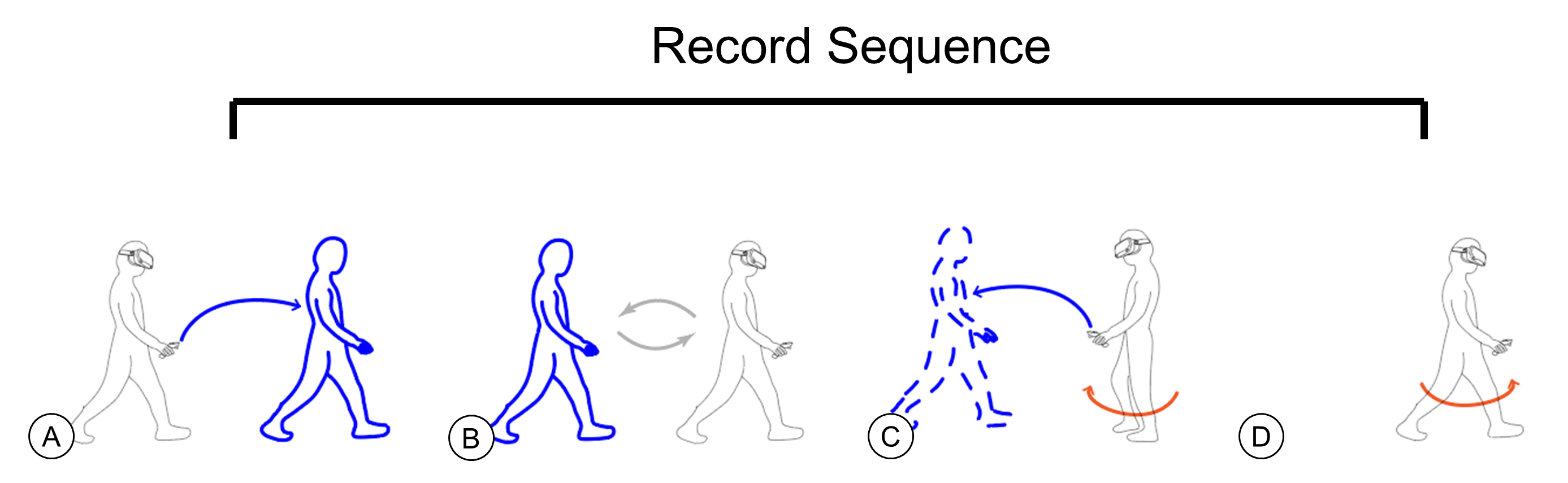}
  \caption{A \textbf{teleportation} technique can be assembled through the spatiotemporal arrangement of \projectName{} with an additional \textbf{remove} technique. A user can record their sequence of actions, including (A) spawning a clone, (B) switching control to the newly spawned clone, (C) turning back and \textbf{remove} the original avatar, and (D) returning to the original orientation. Subsequently, they can easily apply this stored, recorded sequence to themselves whenever they wish to teleport without redoing each decomposed step. This creates an automated process for achieving \textbf{teleportation}.}~\label{fig:auto-teleport}
\end{figure*}

In this subsection, we explain how \projectName{} can create an automaton that acts as a new interaction technique beyond the system's original set of interactions by combining existing interactions spatiotemporally in the VR system. As illustrated in Figure~\ref{fig:auto-teleport}, let's consider a VR environment where the fundamental components of \projectName{} are implemented, along with an additional technique called \textit{remove} (which allows the removal of a group or a single clone).

To establish an automaton enabling a teleportation technique (where users can instantly appear at a different location), the user can initiate the recording function in replay mode (temporal manipulation). They can then spawn a clone in front of themselves, switch control to the newly created clone (spatial manipulation), turn around and \textit{remove} the avatar that has been deactivated, and finally, return to the original forward orientation. After completing this sequence of actions, the user will find themselves at a new location, at which point they can stop the recording.

From that point forward, whenever the user needs to perform a teleportation, they can trigger these automated recordings from the replayed 3D carousel menu, applying them to themselves. This results in a reusable and functional \textit{teleportation} technique. Notably, each step falls within the system's original capabilities, making it highly understandable, as they consist of basic, straightforward spatiotemporal operations.

When a system contains a different interaction set, then \projectName{} may provide the possibility to form a different resulting solution space. For example, if we take \textit{remove} out of the aforementioned VR system and add \textit{scale} function to it (where users are allowed to resize the avatar and clones), then we can possibly create a World-in-Miniature function ~\cite{Stoakley:WIM} with \textbf{synchronous} control between the users and a large giant clone.

\section{Implementation}
\label{sec:implementation}
We implement \projectName{} using Unity version 2020.3.18f1, running on an HP VR Backpack G2 computer equipped with a NVIDIA GeForce RTX 2080 graphics card. 
The VR content is streamed to a Meta Quest 2 headset via Quest Link over a USB cable. 
The display resolution is 1832 × 1920 per eye, and the refresh rate is set to 72 Hz. 
The user holds two Meta Quest 2 controllers as input devices. 

The tracking data from the headset and controllers are distributed to both the original avatar and its clones.
Specifically, we map the positions and the rotations of the headset and the controllers to the corresponding joints (i.e., the neck and the wrists) on a humanoid avatar and use Unity's built-in inverse kinematics system to animate the avatar's full body movements.
Hence, one limitation of our system is that users can only record their upper body movements and can not control their legs to form more dynamic poses.

\subsection{User Interface}
To help users get started faster, we have implemented a 2D menu that can be toggled by pressing the controller button.
The menu is mounted on the user's left hand and provides access to all the commands and functions.

Additionally, \projectName{} offers voice command functionality for experienced users.
The voice recognition system is developed using Unity's built-in keyword recognizer, which can respond to pre-defined keywords.
The microphone of the headset captures the voice.
Users can initiate a voice command anytime as the microphone is always active.
Users can also access a list with all available voice commands inside the menu.
It allows users to seamlessly control \projectName{} faster without breaking their immersion.
In situations where a 2D menu is not feasible, such as when the user's hands are fully occupied with other tasks, voice commands become a valuable alternative.

\subsection{Added Interaction Technique}
While we have covered the essential functions enabling \projectName{} in Section~\ref{sec:Clonemator}, its true potential lies in its capacity to integrate and combine existing interactions that were accommodated within the current virtual world, thereby generating novel, adaptable, and understandable solutions.
A different collection of these added interaction techniques may span a different solution space.
Here, we provide an overview of the added interaction techniques we have incorporated into our current implementation. Note that the selection of these interaction techniques is for illustrative purposes and the following user study, the concept of \projectName{} allows VR designers to determine their own set of added interactions for enabling a different possible solution space for the users to explore.

\begin{itemize}[leftmargin=*]
\item{Snap}: Although \textit{\indirectSpawning{}} provides a quick way to spawn a clone at a distance, it lacks precision in determining the spawning position.
To further ease the user's effort when positioning the clone, \projectName{} introduces two snapping methods, which can be toggled using the controller button.

\textit{Grid Snapping} divides the world into a grid, with each cell having a length equivalent to the user's arm length.
It ensures that the spawned clones can interact with each other if they are located in two adjacent cells.
This method is convenient for maintaining consistent offsets between avatars.

\textit{Nearest Object Snapping} snaps the clone to the closest object while maintaining an arm's length offset.
This method is useful when spawned clones need to interact with a specific object.
Both snapping methods can be customized by using a slider to adjust the offsets according to different situations.

\item{Avatar Rotation and Teleportation}:We enable users to navigate easily by rotating their bodies or teleporting using the controller.
Since \projectName{} currently synchronizes physical movements, users can use avatar rotation and teleportation to prevent unintentional movements of synchronous clones.
For example, when users rotate themselves using the controller, the clone will remain stationary, enabling them to establish a rotational offset.

\item{Scale}: Users can adjust the size of clones by pressing the thumbstick backward and forward during \textit{\indirectSpawning{}} or when grabbing an existing clone.
Users can also switch between clones of different sizes to perform interactions at varying scales~\cite{Izumihara:Transfantome}.


\item{Mirror}: Mirroring allows the user to flip a clone's movement horizontally, creating a mirror image of their own actions.
For instance, when the user raises their right hand, the clone will raise its left hand, and vice versa.
It provides intuitive control, especially when the user and the clone face each other as their movements are reflected, much like looking into a mirror.
Mirroring is also helpful for activities that require symmetric movements, such as swinging a long jump rope together or performing synchronized dance moves.

\item{Remove}: Users can select and remove a clone using a ray.
If the clone is holding an object, the object will dissolve and appear near the user.
The mental model behind this feature is that the removed clone will merge back with the user.
For example, the user can assign a clone to gather wood and then remove the clone to obtain the collected wood. 

\item{Undo}:
\projectName{} also provides an Undo feature, which allows the user to cancel previous spawning commands.
It is especially useful when the user wants to remove multiple clones created by \textit{Auto Spawning} at the same time.
Our system also allows users to undo grouping and duplication commands, reducing manual effort.
\end{itemize}

\section{Preliminary User Study}
Since \projectName{} can possibly utilize infinite supporting interaction techniques to form an even larger solution space, it was challenging to identify a particular collection of supporting interactions to show its full potential.
As the first step to understanding \projectName{}, we conducted a preliminary study whose main evaluation is around \projectName{}'s key components only, including user experience and flexibility in solution-making. To allow participants to focus on the key components of \projectName{}, we utilized the smaller supporting interaction set employed in our walkthrough and implementation, and we restricted the replay function to record only the user's physical actions.


\subsection{Design, Tasks, and Participants}

The study includes 9 tasks in the VR camping scene presented earlier in Section~\ref{sec:walkthrough}, and is composed of two sessions: a first session with example solutions and a second session without. 

In the first session, participants completed five tasks, including four specific tasks from the walkthrough in Section~\ref{sec:walkthrough} (hammering multiple pegs, catching fish, cooking while adding ingredients, and fanning the campfire while cutting beef) and a free-exploration task. 
Prior to the five tasks, a video tutorial was given, explaining the tasks' goal, providing the sample solutions from the walkthrough, and demonstrating the necessary UI elements and actions. 
Participants had to replicate the sample solutions in the VR scene after viewing the video. 
The free-exploration task lets participants test the operation not covered in the first four tasks, such as remove, duplicate and group, and allowed participants to explore the scene freely with all operations available. 
The first session familiarized participants with the system and provided a general understanding of \projectName{}'s experience. \cy{A systematic overview of these tasks might be needed. For now, they seem to be very random.}

In the second session, participants had to complete the other four tasks, including:
\begin{itemize}[leftmargin=*]
    \item Moving a table that is too heavy for a single person, so the participants have to collaborate with or utilize multiple clones
    \item Passing multiple basketballs as fast as possible from a starting point to a target that is 9 meters away without teleportation
    \item Fetching a slice of beef at 2.5 meters height from the top of a van
    \item Fetching an apple at 7.5 meters height from the top of a tree, where a distinct solution from fetching beef was required.
\end{itemize}
Participants were only given the task goal and requirement without any further instructions provided. 
These tasks are designed for utilizing more complex clone manipulation, and since participants had to come up with their own solutions, this session is aimed at testing not only the user experience but also the richness of the solutions that \projectName{} can support.

\begin{figure*}[!h]
\centering
   \includegraphics[width=\textwidth]{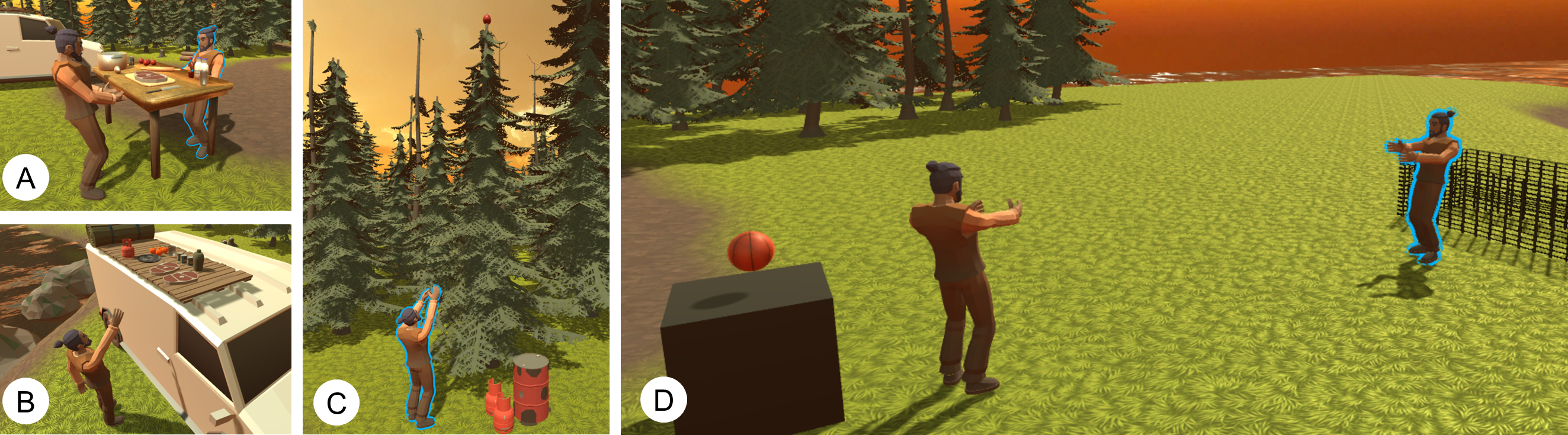}
  \caption{Study tasks without sample solutions. Participants have to complete four challenges without explicit instructions. These challenges included: (A) lifting a heavy table, (B) fetching a slice of beef from the top of a van, (C) fetching an apple from the top of a tree using a solution different from that used for fetching the slice of beef, and (D) passing basketballs across a distance without teleportation as fast as possible.}~\label{fig:study_tasks}
\end{figure*}

We recruited 12 participants (8 male, 4 female) who had no prior knowledge of \projectName{}'s details, aged from 19 to 27 (\mean{23.8}, \sd{2.4}) through word of mouth. 
2 had VR experience more than once a week, another 3 had VR experience about once a month, 1 had VR experience about once every three months, and the remaining 6 had very rare VR experience (about once a year or less). 
The apparatus used for the study is the same as mentioned earlier in Section~\ref{sec:implementation}.




\subsection{Solutions Generated by Participants}
\begin{figure*}[!h]
\centering
   \includegraphics[width=\textwidth]{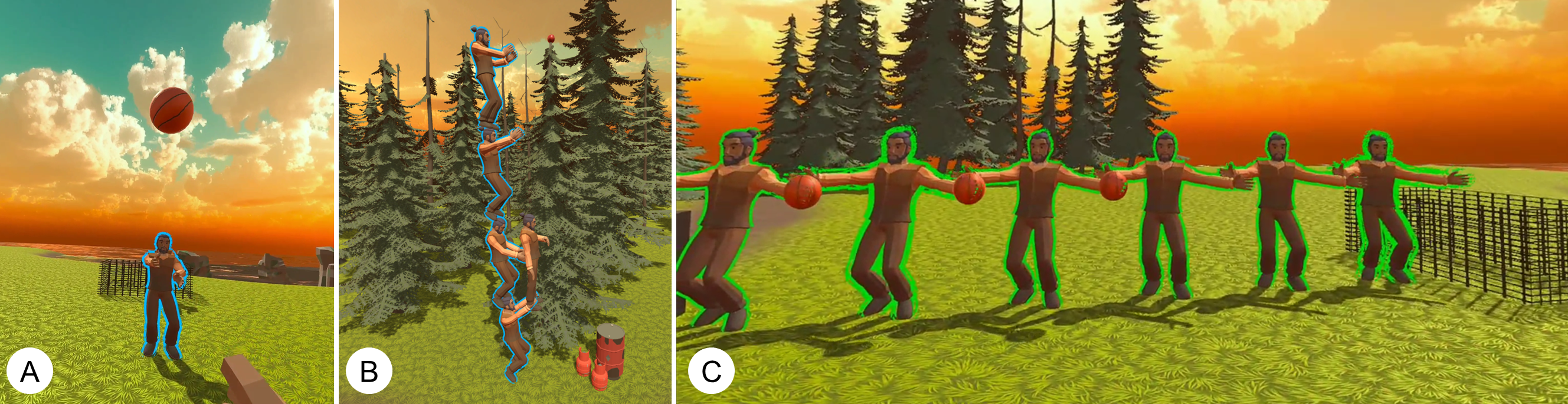}
  \caption{Solutions come up by participants: (A) The user throws a ball and catches it with a synchronous clone. (B) The user replays the lifting movement to multiple clones stacked vertically, forming an elevator. (C) A bucket brigade formed by a chain of clones with synchronous movement.}~\label{fig:example_solutions} 
\end{figure*}

During the second session of the study, we observed that participants were able to come up with different solutions for the given tasks.
For example, P6 used a synchronous clone to catch a basketball he threw (Figure~\ref{fig:example_solutions}A), while P5 and P11 recorded themselves throwing a ball and applied the recording to a clone, catching the ball at the destination by themselves.
P2 successfully built a bucket brigade made of clones using the \textit{\directSpawning{}} technique and set their interaction modes to synchronous (Figure~\ref{fig:example_solutions}C).
The first clone grabbed a ball with its right hand and passed it to its left hand, while the next clone used its right hand to grab the ball held by the left hand of the previous clone, forming a pipeline to pass multiple balls efficiently.

For fetching a high object (including fetching beef and fetching an apple), P4 used \textit{Relative Spawning} to spawn a clone in front of the apple by setting the reference object to the gas cylinder next to her and then switched to the clone to get the apple.
P3, P6, P7 and P11 directly threw a clone to the top of the tree and switched to it.
P5 recorded the action of grabbing and lifting objects and applied it to vertically stacked clones with different phase shifts, forming an elevator that allowed the clones to pass him to the tree top (Figure~\ref{fig:example_solutions}B).
P8 created a step stool by positioning a static clone near the camper van and used it as a platform to reach for the beef.

Overall, participants created 5 different solutions for fetching a high object and 4 different solutions for passing basketballs, which indicate ~\projectName{}'s flexibility in providing dynamic, adaptable solutions.
\cy{maybe a short overview like: x different solutions were executed, should be mentioned.}

\cy{completion time, number of clones generated, clone removal, and error rate (\# of removal / \# of clones generated)}

\subsubsection{Overall Experience}
On \pquote{the overall experience of using the system is enjoyable,} participants reported an average of 6 points (\sd{1.13}) on a 7-point Likert scale (1-\pquote{strongly disagree}, 7-\pquote{strongly agree}).
Participants responded positively about the experience of completing tasks with \projectName{}, including
\pquote{It's pretty cool. It's interesting and fun.}{P1}, \pquote{It's a fun game.}{P2, P7, P10}, \pquote{It's pretty fun, even though it's my first time using it.}{P3}, \pquote{It's quite fun. The experience is interesting.}{P5}, 
\pquote{You can do many things due to variability and combinations, so it's pretty interesting.}{P11}. \cy{These comments are kinda repeated, another alternative is merging them with the same keywords e.g. fun., but I'm also okay with throwing an overwhelming amount of positive quotes to readers.}

\subsubsection{Understandability and Usefulness}
Regarding the system's usefulness, participants noted \pquote{Even without instructions, one should be able to figure it [the solution] out.}{P1} and \pquote{I can think of a solution easily.}{P12}. For learning the overall control of the system, participants added \pquote{It is not difficult to learn, there are many things but all the operations are basic.}{P2}. \pquote{Once you're familiar with what you can do with the clones, it's quite simple.}{P11}

 Regarding spawning the clones, participants express it's natural as \pquote{The method of creating clones is quite intuitive.}{P3}. Specifically, \pquote{Out-of-body is convenient and most intuitive}{P1, P12} ... \pquote{because one can determine the position of their clone by his body.}{P1}, and provides their view on tradeoff between different spawning methods as \pquote{It's harder to image its result for auto-spawing but it has huge potential for mass deployment.}{P12} In terms of controlling the clones, \pquote{simultaneously waving the wind and cutting foods [using synchronous mode] is intuitive because the movements are consistent.}{P1}, and P3 mentioned her controlling metaphor, by \pquote{to think about how to operate different bodies at the same time, it's like controlling a machine remotely.}{P3} 


\subsubsection{Use Cases and Further Applications}
The participants have also suggested additional features. For example, editing or managing replayed clones \pquote{Replay could benefit from naming, searching, and categorizing. Also, I hope to extract certain actions during a replay sequence}{P7}, as well as more dexterous control with \pquote{complex actions such as finger movement.}{P11}

Although we make sure the keyword or notion of automation is never mentioned in our study instruction, P2 drew a comparison between \projectName{} and such concept, by \pquote{It [the system] is like writing code modules and microservices. Using the switch or replay to trigger pre-written functions is like a game plugin.} and \pquote{similar to a macro or script but for repeated tasks in VR.}

Furthermore, participants proposed additional potential applications of \projectName{}, which encompassed \pquote{3D modeling in VR}{P1},  \pquote{gaming}{P1, P2, P3, P5, P8, P10, P11, P12}, and \pquote{demonstration or teaching}{P6, P7, P12}. Specifically, P7 suggested that \pquote{ [\projectName{}] can be used for teaching and even demonstrate tasks that require collaboration or teamwork but by one's self.}{P7}, while participant P2 mentioned the composition of clones \pquote{can improve the experience of games like Story of Seasons, which involves a lot of repeated farming tasks.}
\subsection{Study Summary}
The study results indicate that \projectName{} provides diverse and adaptable functionalities for various tasks, and the key operations of \projectName{} deliver a user-friendly and enjoyable experience with intuitive controls.

\ys{We did not ask the participants to finish the tasks as soon as possible. P8 said that he didn't want to apply quick but boring solutions during the execution.}


\section{Related Work}
The presented work draws on several areas of previous research, including Replicating Movement in the Real World, Beyond-Real Interaction, and Programming by Demonstration.

\subsection{Replicating Movement in the Real World}
Previous studies in teleoperation have investigated the potential of replicating control or movement in real-world scenarios. 
This involves remotely controlling one or multiple robots by one single user. 
For instance, Glas et al. ~\cite{Glas:TeleoperationOfMultipleSocialRobots} designed a coordination system that enables a user to operate four robots at the same time. Researchers have also explored ways of controlling two robotic arms synchronously or asynchronously through the use of EMG signals and gaze control~\cite{Nakanishi:SyncAndAsyncControlOfRobots}. 
Additionally, in Transfantome ~\cite{Izumihara:Transfantome}, users can control two robots concurrently at different scales by embodying two proxy bodies that represent each robot. Takada \ea ~\cite{Takada:ParallelPingPong} utilized parallel embodiment to allow users to simultaneously operate two robot arms to play ping-pong against two opponents. These studies demonstrate the potential of duplicating control and movement in enhancing physical operation. 

However, these methods are restricted by real-world hardware resources, leading to limitations in the number of clones or replications. 
In \projectName{}, we systematically explore such a concept in virtual reality environments, allowing users to operate an unlimited number of clones of varying sizes, timing, and replayability, thereby fully realizing the potential for movement cloning and replication.

\subsection{Beyond-Real Interaction in VR}
VR offers a unique platform for exploring interactions that are impossible to experience in the physical world. Such interactions have been coined as "Beyond-real interaction"~\cite{abtahi_beyondreal_2022} by Parastoo \ea Researchers have investigated a variety of beyond-real interactions, including dynamically changing arm length for reaching objects~\cite{ivan_gogointeraction_1996}, scaling the body for navigation~\cite{abtahi_imagiant_2019}, and using spherical proxies to interact with distant objects~\cite{pohl_poros_2021}.

Researchers have also begun exploring the possibility of multi-embodiment or duplicating body parts in VR. Studies have shown that users can adapt to the six-digit hand while maintaining a sense of ownership and agency with it~\cite{Hoyet:SixFingers}. Similar results can also be applied to having a third arm~\cite{Drogemuller:RemappingAThirdArm} and even multiple bodies~\cite{Miura:MultiSoma}. Furthermore, users can develop an implicit dual motor adaptation when switching between two virtual bodies~\cite{Verhulst:ParallelAdaption}.

Beyond the experience of having supernumerary body parts, researchers have also investigated the resulting changes in performance. For example, it has been shown that users can increase task efficiency and reduce physical movement by leveraging additional hands~\cite{Schjerlund:NinjaHands} or bodies~\cite{Schjerlund:OVRlap}.
Smith \ea have also demonstrated that users can coordinate two bodies simultaneously and develop different strategies for handing off a cube~\cite{DualBodyBimanualCoordination}.
However, while controlling multiple clones, sharing multiple perspectives (\ie more than two) simultaneously could also increase control complexity, thus adding task completion time~\cite{Miura:MultiSoma}.

\projectName{} is situated within the ``duplication'' category of Beyond-Real Interaction's taxonomy~\cite{abtahi_beyondreal_2022}. Compared to previous work, our research further explores the interaction between the user and their clones, leveraging the ability to create clones with various temporal and spatial properties to give users greater control. This approach builds upon previous research and takes it to the next level by enabling the possibilities for user-clone interactions.

\subsection{Programming by Demonstration}
\cy{or, Programming by Example}

Previous research has investigated Programming-by-Demonstration (PbD)~\cite{lieberman2001your} to enhance the accessibility of end-user programming. PbD liberates users from the constraints of dedicated programming languages or scripts by enabling them to create their automated solutions by demonstrating "an example of what they wanted it to do" directly to the system~\cite{lieberman2001your}.

For instance, Sikuli~\cite{yeh_sikuli_2009} allows users to utilize screen-captured elements as their script and search input, bypassing the need for expert knowledge of the parameters or names of the corresponding UI. Additionally, SikuliBot~\cite{kim_sikulibot_2014} extends the same concept to physical interfaces by enabling users to take real-world images of physical buttons or touchscreen positions, and a robotic actuator automatically operates these physical UI elements based on the users' arrangement of the images. In addition to images or screenshots, recording and replaying a sequence of users' interactions further enhances the automation process with PbD. For instance, Ringer~\cite{barman_ringer_2016} allows non-coders to record a sequence of their website interaction, which generates a script that interacts with the page as demonstrated by the users. Similarly, Mau'{e}s \ea~\cite{maues_keepdoing_2013} enables users to perform a smartphone task that they would like to automate, and their system generates the automation based on these latest actions. Moreover, SUGILITE~\cite{li_2017_suglite} combines voice input and app demonstration to enhance the automation pipeline for non-programmers.

While these approaches have lowered the barrier to automation for desktop, web, and mobile applications, \projectName{} explores this concept in the context of virtual reality. Specifically, \projectName{} builds upon the PbD concept for task automation by allowing users to record and replay their VR demonstration, which can be rearranged, distributed, and composited with each other in the form of clones in VR. Users can create and control clones to perform specific actions as if they were controlling their own bodies, without requiring dedicated crafting of different interaction techniques for different situations.


\section{Discussion and Future Work}\label{sec:discussion}

\subsection{Snapping and Alignment}
Composing multiple clones to create custom automators often requires precise alignment of the clones.
\projectName{} addresses this by offering several alignment techniques, such as \textit{Auto Spawning} and \textit{Relative Spawning}, as well as \textit{Grid Snapping} and \textit{Nearest Object Snapping} modes when spawning a clone. 

While these techniques can handle situations in our case, more advanced techniques are necessary for complex scenarios, such as when the user wants to spawn a clone in front of an object that is significantly smaller than the object in front of the user since the required offsets may not be the same.
This can lead to misalignment, making it difficult for the clone to interact with the smaller object simultaneously.


\cy{Possible solutions (1) manual adjustment (2) visualize forward vector (3) preview clone location}

\subsection{Cloning with objects in hands \vs{} without}
One confusion we observed during the study and our own implementation is whether the object held in hand should be cloned when spawning a user's clone. If not, then interactions such as hammering multiple pegs simultaneously can't be easily achieved. However, if we clone the held objects every time, excessive objects would be produced unnecessarily.

We believe this problem essentially pertains to defining the boundary of a clone's ownership. Since we currently see the need for both alternatives, we suggest that future systems allow users to adjust what they want to include in the cloning and what they don't. While our implementation currently achieves this by setting different object cloning modes for different spawning methods (\ie{} while auto and relative spawn clone the object in hands, the rest don't), we see a complete decoupling between object cloning and spawning method as a better future option.




\subsection{Further Supporting Techniques}
With only the key components of \projectName{}, the system can possibly play with any interaction techniques and enable a bigger solution space. However, we have identified several supporting interaction techniques closely related to the concept of \projectName{} itself, which we believe is not necessary to enable \projectName{} but will improve the automation process.
We list these interaction features here to inspire future research on the system.
\begin{itemize}[leftmargin=*]
\item\textbf{Temporal Transformation:}
While we have incorporated interaction techniques about spatial transformations (\eg{} scaling a clone or group) in our current implementation, the comparable future supporting technique could be temporal editing of clones. This could potentially involve trimming unnecessary replayed segments, extracting replayed clips to create new  sequences, and implementing fast-forward or rewind functionalities within a replay.



\item\textbf{Visualization and Mesh Control:}
While we currently present the user avatar and clones with full-body mesh, visualizing only the upper body of the avatar or even just the hands could reduce the visual burden when many automators are working at the same time (\eg{} utilizing multiple clones to create a pantograph mechanism for achieving adjusted CD gain in fetching). 
Additionally, allowing users to change the coloring of clones would make it easier for them to manage them using color labels systematically. 
Moreover, changing the mesh or texture of a clone could enable a customized form of automators. 
For instance, when using a string of static clones to build a row of fences, giving them a wooden texture or simplified mesh would be more realistic and aesthetically pleasing.


\item\textbf{Shortcut Bindings:} In our existing implementation, to reuse a previous recording, users must initially access a carousel menu. To duplicate a group of clones, they need to target an existing group and execute the duplication.

While this design sufficiently facilitates automated tasks, an enhanced interaction could enable users to bind the reuse or duplication of specific actions or groups to a button shortcut. This would prove especially convenient for frequently used interaction techniques.

\end{itemize}






\subsection{Merging a Clone's Experience}
\begin{figure*}[!h]
\centering
   \includegraphics[width=\textwidth]{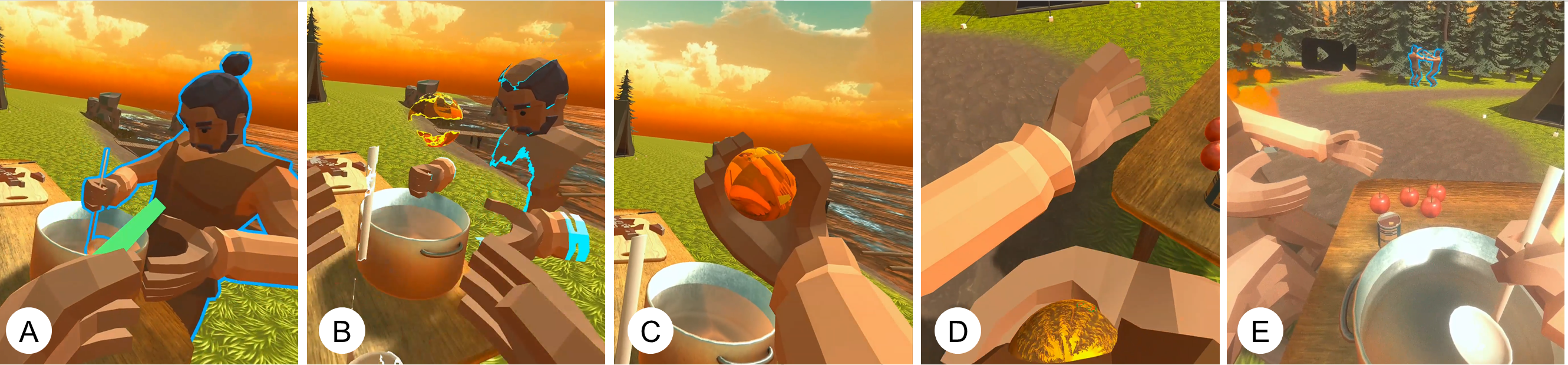}
  \caption{A proof-of-concept feature allowing users to relive a clone's experience: (A) The user removes a clone and (B)-(C) a memory orb appears. (D) The user places the memory orb into his chest and (E) experiences the clone's memory.}~\label{fig:memory_orb}
\end{figure*}
The Shadow Clone Technique heavily inspires us in the popular anime, Naruto.
One of the key features is the user's ability to gain the clones' experiences when they disperse.
In \projectName{}, we have implemented a similar feature where a memory orb, inspired by the film \textit{Inside Out}, appears in front of the user upon removing a clone.
By grabbing the memory orb and placing it into his chest, the user can relive the clone's experience by watching a replay captured by the clone's eye (Figure~\ref{fig:memory_orb}).

However, a question remains: can the user experience a clone's sensations other than sight?
For example, imagine removing a clone that has been stirring a pot for half an hour and instantly feeling the soreness in their arm.
This presents an interesting design space for future exploration.






\subsection{Beyond Problem Solving}
Although, for our VR study, we implemented and demonstrated \projectName{} with its problem-solving capacity, we believe the work has further potential to be utilized in other areas.

To begin with such extension, we believe \projectName{} can serve greatly for social scenarios such as VRChat~\cite{vrchat} or BeanVR~\cite{beanvr}. With \projectName{}, in social platforms, players can compose versatile body motions with single or multiple replayed clones, forming groups of clones with intriguing postures as signs, or deploying multiple dynamic dancing clones as social gestures such as Figure ~\ref{fig:dancing_drawing_sign}. This extension also hints at the potential for ~\projectName{} to function as a creative system where individual users can harness the cloning operations to generate diverse artistic expressions, including forms that traditionally may require multiple users, such as movies (multiple actors/ actress) and dances (multiple dancers).

Another prospective area is a multi-robot system. While our system is conceptually similar to the "Distributed Autonomous System"~\cite{districuteautonomous} that utilizes multiple independent robot units to work coordinately and achieve a more complex task, it has the potential to serve as a control mechanism for future swarm user interfaces~\cite{Zooids} or multiple robotic arms.
By leveraging the flexibility of \projectName{}, users can control multiple elements as intuitively as controlling their clones.
For example, when a user spawns a clone, a corresponding robot will appear in the real world at the same position.
It can also facilitate users in switching between multiple robots~\cite{Izumihara:Transfantome} or recording and replaying actions for teleoperation~\cite{Mimic}.

Besides, previous video games have already leveraged different aspects of clone-based mechanisms to create novel gaming experiences.
For example, Quantum League~\cite{QuantumLeague} and Time Rifters~\cite{TimeRifters} allow players to team up with their past selves in a shooting game.
More recently, The Last Clockwinder~\cite{TheLastClockwinder} enables players to create clones that replay their actions and construct a pipeline to maximize the resource harvesting throughput. 
While \projectName{} systematically extends the possibilities of interactions between clones by providing multiple interaction modes and showcasing how they can be combined with supporting techniques for even more complicated tasks, we believe our work will inspire game designers to explore and develop more engaging and innovative gaming mechanisms.

\section{Conclusion}
We have presented \projectName{}, a VR system that allows a user to create and collaborate with clones to accomplish complex tasks. We have demonstrated the potential of \projectName{} by showcasing concrete examples based on our systematic exploration of spawning clones, clones' properties, and interactions with traditional techniques. We also have shown from our preliminary study that participants were able to intuitively, creatively, and also enjoyably use \projectName{}. With \projectName{}, we see the user's avatar could be the generic automator waiting to be dispatched.

\begin{acks}
\end{acks}

\bibliographystyle{ACM-Reference-Format}
\bibliography{shadowclone}


\end{document}